\documentclass[prb,aps,amssymb,twocolumn,showpacs,amsmath]{revtex4}

\usepackage{graphicx}

\begin{document}

\title{Spin-Photon Dynamics of Quantum Dots in Two-mode Cavities}
\author{Florian Meier}
\email{meier@physics.ucsb.edu}
\author{David~D. Awschalom}
\email{awsch@physics.ucsb.edu} \affiliation{Center for Spintronics
and Quantum Computation, University of California, Santa Barbara,
California 93106, USA }
\date{\today}

\begin{abstract}
A quantum dot interacting with two resonant cavity modes is
described by a two-mode Jaynes-Cummings model. Depending on the
quantum dot energy level scheme, the interaction of a singly doped
quantum dot with a cavity photon generates entanglement of
electron spin and cavity states or allows one to implement a
\textsc{swap} gate for spin and photon states. An undoped quantum
dot in the same structure generates pairs of polarization
entangled photons from an initial photon product state. For
realistic cavity loss rates, the fidelity of these operations is
of order $80$\%.
\end{abstract}


\maketitle

\section{Introduction}
\label{sec:intro}

The electron spin in quantum dots (QD's) is among the most
promising candidates for quantum information processing in the
solid state.~\cite{loss:98,wolf:01} Optical selection rules make
it possible to control and measure spins in QD's
optically.~\cite{imamoglu:99,imamoglu:00,pazy:03,chen:04} For
pairs of QD's embedded in a cavity, in the strong-coupling limit
cavity photons can mediate an effective exchange interaction
between electron spins.~\cite{imamoglu:99,imamoglu:00} The Faraday
rotation of a single photon interacting with an off-resonant QD
has recently been discussed for the implementation of Bennet's
quantum teleportation scheme and the generation of spin-photon
entanglement.~\cite{leuenberger:04} Because the coupling of cavity
photons to an off-resonant QD is weak, such schemes require long
electron spin decoherence times, a high cavity $Q$-factor, and
control of the cavity $Q$-factor on a picosecond time-scale.

\begin{figure}
\centerline{\mbox{\includegraphics[width=8cm]{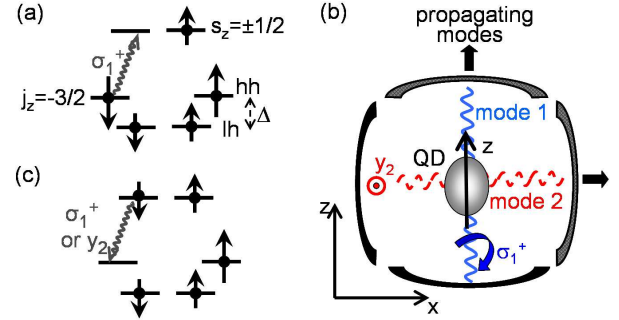}}}
\caption{(a) Characteristic level scheme of, e.g.,  a CdSe
nanocrystal. The crystal anisotropy leads to a splitting $\Delta$
of hh ($|j_z|=3/2$) and lh ($|j_z|=1/2$) states. (b) Schematic
representation of the cavity-QD system. The circularly polarized
mode $|\sigma_1^+\rangle$ propagating along direction $1$ (aligned
with the QD anisotropy axis $z$) and the linearly polarized mode
$|y_2\rangle$ propagating along $2$ are resonant with the hh-trion
transition. (c) The trion state can decay by emission of a photon
into state $|\sigma_1^+\rangle$ or $|y_2\rangle$.}\label{Fig1}
\end{figure}

Recent progress in microcavity design has led to mode volumes
close to the theoretical limit $(\lambda/n)^3$ and $Q$-factors of
order $5\times10^3$, approaching the strong-coupling limit for QD
cavity-QED.~\cite{kiraz:03,vuckovic:03} A QD coupled to one
circularly polarized cavity mode is described by the
Jaynes-Cummings model~\cite{puri:01} and is expected to show
phenomena such as vacuum Rabi oscillations. Here, we theoretically
study the coherent dynamics of a QD coupled to two cavity modes
$1$ and $2$ with different spatial distribution and polarization
[schematically shown in Fig.~\ref{Fig1}(b) for orthogonal
propagation directions]. The design of a cavity with small mode
volume and two degenerate, orthogonal modes with circular and
linear polarization at the site of the QD is difficult, but
possible in principle (see Sec.~\ref{sec:exp} below). The aim of
this paper is to show that such a system has interesting
applications as interface between electron spins and photons
because the second cavity mode gives rise to intriguing effects.
Most notably, photon transfer between the cavity modes via an
intermediate trion state is controlled by the spin state of the
QD, opening a wide range of possible applications. We show that
(i) for cavity modes in resonance with the heavy hole (hh)-trion
transition, entanglement of the electron spin and the cavity
modes, i.e., the photon \emph{propagation direction} is generated.
(ii) For cavity modes in resonance with the light hole (lh)-trion
transition, the strong-coupling dynamics can be used to implement
a \textsc{swap} of spin and photon states, an operation which
would allow one to transport a spin quantum state over large
distances.~\cite{park:70} The quantum state of the photon is
encoded in the occupation amplitudes of the two cavity modes.
Hence, the system discussed here provides a natural interface
between spins and linear-optics quantum information
schemes.~\cite{leuenberger:04,knill:01} For cavities with
switchable $Q$-factors, the fidelity of all operations, $1-{\cal
O}(g/\Delta) \simeq 1$, is limited only by off-resonant
transitions, where $g$ is the coupling constant for the trion
transition and $\Delta$ the hh-lh splitting. However, even for
lossy cavities without time-dependent control parameters, the
fidelity is of order $80$\% for realistic cavity loss rates. We
also show that (iii) an undoped QD efficiently generates pairs of
entangled photons from initial photon product states.

We consider a QD with an anisotropy axis $z$ determined by crystal
or shape anisotropy which leads to a splitting $\Delta$ of hh and
lh states at the $\Gamma$ point (Fig.~\ref{Fig1}). The ground
state of a singly doped QD is determined by the spin of the excess
electron, $\alpha |\!\uparrow\rangle + \beta
|\!\downarrow\rangle$. In the following, we evaluate the dynamics
of the QD--cavity system after injection of a photon in state
$|\sigma_1^+\rangle$ at $t=0$. For quantitative estimates, we
consider CdSe nanocrystals and adopt the model of
Ref.~\onlinecite{efros:92} where the anisotropy is treated
perturbatively in the quasi-cubic approximation. The coupling
constant $g$ for a photon with polarization vector ${\bf e}$
resonant with the hh (lh)-trion transition is determined by the
interband matrix element of the momentum operator, ${\bf e}\cdot
\hat{\bf p}$, and the overlap integral of the $1S_e$ and
$1S_{3/2}$ ($1S_{1/2}$) electron and hh (lh) wave functions. In
addition to the strong-coupling criterion that $g/\hbar$ be large
compared to the QD spontaneous emission rate and the cavity loss
rate, we also assume that $g/\hbar$ is large compared to the hole
spin relaxation rate.

In the following, we show that systems such as the one shown in
Fig.~\ref{Fig1}(b) allow one to generate entanglement between an
electron spin and the cavity state (Sec.~\ref{sec:entanglement}),
to implement a spin-photon \textsc{swap} gate
(Sec.~\ref{sec:swap}), and to efficiently generate pairs of
polarization-entangled photons (Sec.~\ref{sec:photons}). In
Sec.~\ref{sec:exp}, we discuss how a microcavity with the mode
structure shown in Fig.~\ref{Fig1}(b) can be engineered and
illustrate that the implementation of the schemes discussed in
Secs.~\ref{sec:entanglement}, \ref{sec:swap}, and
\ref{sec:photons} is feasible for microcavities with $Q$-factors
exceeding $10^4$.

\section{Spin-photon entanglement}
\label{sec:entanglement}

The interaction of a QD with a hh valence band ground state
[Fig.~\ref{Fig1}(a)] with the circularly polarized cavity mode
propagating along $1$, $|\sigma_1^+\rangle$, and the linearly
polarized cavity mode with polarization vector ${\bf e}_y$
propagating along $2$, $|y_2\rangle$, is described by a two-mode
Jaynes-Cummings model. A photon injected into $|\sigma_1^+\rangle$
at $t=0$ induces transitions from $| \!\uparrow\rangle$ to the
trion state $|X^- \rangle = \hat{c}^\dagger_+ \hat{c}^\dagger_-
\hat{h}_{-}|G\rangle$, where $|G\rangle$ is the ground state of
the QD without excess charge and $\hat{c}_\pm$ ($\hat{h}_\pm$) the
electron annihilation operator for the $1S_e$ conduction band
level with $s_z = \pm 1/2$ (the $1S_{ 3/2}$ hh level with $j_z=\pm
3/2$). The trion state $|X^- \rangle$ has two possible decay paths
via emission of a photon in state $|\sigma_1^+\rangle$ {\it or}
$|y_2\rangle$ [Fig.~\ref{Fig1}(c)]. In both cases, the QD spin
remains unaltered by the cycle of photon absorption and subsequent
emission because spin-flip transitions involving the lh-component
are dipole forbidden within the model of
Ref.~\onlinecite{efros:92}. The interaction of QD and cavity modes
is
\begin{eqnarray}
    \hat{H}_{\rm I} &= & g_1 \left( \hat{a}_1 \hat{c}^\dagger_-
\hat{h}_{-} + h.c. \right) \nonumber \\ && + g_2 \left[ \hat{a}_2
\left( \hat{c}^\dagger_- \hat{h}_{-} + \hat{c}^\dagger_+
\hat{h}_{+} \right) + h.c. \right], \label{eq:interact-h}
\end{eqnarray}
where $\hat{a}_{1}$ ($\hat{a}_{2}$) is the photon annihilation
operator for mode $|\sigma_1^+\rangle$ ($|y_2\rangle$) and $g_1$
($g_2$) the corresponding coupling constant. The free Hamiltonian
$\hat{H}_0= \delta \left( \hat{a}^\dagger_1 \hat{a}_1 +
\hat{a}^\dagger_2 \hat{a}_2 \right)$ is determined by the detuning
$\delta $ between the photon frequency $\omega$ and the trion
transition energy.

\begin{figure}
\centerline{\mbox{\includegraphics[width=9.3cm]{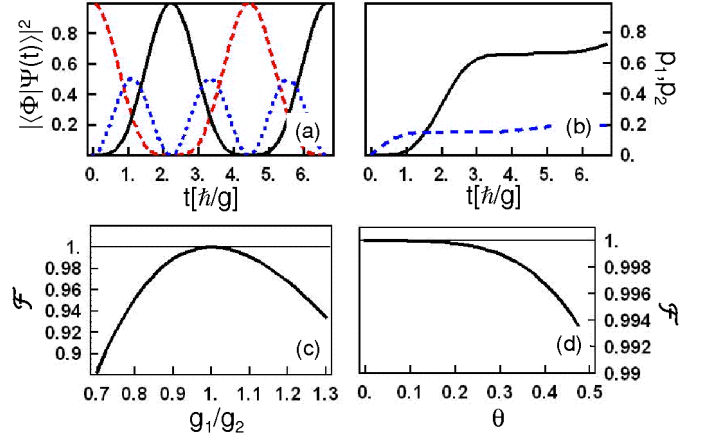}}}
\caption{(a) Time evolution of \mbox{$|\Psi(0) \rangle=
|\!\uparrow;\sigma_1^+\rangle$}. The probabilities $|\langle
\uparrow;\sigma_1^+|\Psi(t) \rangle|^2$ (dashed), $|\langle
\uparrow; y_2|\Psi(t) \rangle|^2$ (solid), and $|\langle X^-;
0|\Psi(t) \rangle|^2$ (dotted) are shown as a function of time.
(b) Probability for photon detection outside the cavity in
direction $2$ (solid) and $1$ (dashed) obtained from numerical
integration of Eq.~(\ref{eq:mastereq}) for $\kappa_1=0.2g/\hbar$,
$\kappa_2=g/\hbar$, and $\hat{\rho} (0) = | \! \uparrow;
\sigma_1^+\rangle \langle \uparrow; \sigma_1^+|$. (c) Fidelity of
spin-photon entanglement generation for $g_1 \neq g_2$. (d)
Fidelity of spin-photon entanglement generation as a function of
QD misalignment.}\label{Fig2}
\end{figure}

While $|\! \downarrow; \sigma_1^+\rangle$ is an energy eigenstate
because of Pauli blocking, the QD state $|\!\uparrow\rangle$ is
coupled to both cavity modes. The time evolution governed by
$\hat{H} = \hat{H}_0 + \hat{H}_I$ leads to transitions from an
initial state $|\!\uparrow; \sigma_1^+\rangle$ to $|\!\uparrow;
y_2\rangle$ via the trion state $|X^-;0\rangle$, where $|0\rangle$
is the photon vacuum. Because the dynamics are controlled by the
QD spin, photon absorption and re-emission leads to entanglement
of the electron spin and the photon cavity mode. This effect is
maximal for $g_1=g_2=g$ and $\delta = 0$, where~\cite{rem1}
\begin{equation}
\hat{H} = \ g | X^-;0\rangle  \left( \langle \uparrow;\sigma^+_1|
+ \langle \uparrow;  y_2| \right) + h.c. \label{eq:ham}
\end{equation}
The initial state $|\Psi(0)\rangle = \alpha |\!
\uparrow;\sigma_1^+\rangle + \beta
|\!\downarrow;\sigma_1^+\rangle$ evolves to
\begin{eqnarray}
|\Psi(t) \rangle &=& \alpha \left[ \cos^2 \left( E t/2 \hbar
\right)| \!\uparrow ; \sigma_1^+ \rangle - \sin^2\left( E t/2
\hbar \right)| \!\uparrow ; y_2 \rangle \right. \nonumber
\\ && \, \, \left.  -   (i/\sqrt{2}) \sin\left( E t/ \hbar \right)|\!
X^- ;0\rangle \right] + \beta |\! \downarrow ; \sigma_1^+ \rangle,
\label{eq:t-evol}
\end{eqnarray}
where $E=\sqrt{2}g$ [Fig.~\ref{Fig2}(a)]. At times $t_n =
(2n+1)h/\sqrt{8}g$, $n$ integer,
\begin{equation}
\alpha |\! \uparrow;\sigma_1^+\rangle + \beta
|\!\downarrow;\sigma_1^+\rangle \rightarrow  |\Psi(t_n)\rangle = -
\alpha |\uparrow; y_2\rangle + \beta |\downarrow;
\sigma_1^+\rangle. \label{eq:ent}
\end{equation}
This demonstrates that, similarly to atom-photon
entanglement,~\cite{raimond:01,blinov:04,sun:04} spin-photon
entangled states of the form of  Eq.~(\ref{eq:ent}) can be
obtained in QD cavity-QED. Alternative schemes for the generation
of spin-photon entanglement have been discussed in
Refs.~\onlinecite{leuenberger:04} and \onlinecite{liu:04}.

According to Eq.~(\ref{eq:t-evol}), the spin-photon entangled
state periodically evolves back into the original product state.
In order to maintain the state $|\Psi_{e}\rangle = -\alpha |\!
\uparrow; y_2\rangle + \beta |\! \downarrow; \sigma_1^+\rangle$
the photon must be extracted from the cavity. In principle, this
is possible by a sudden increase of the cavity loss rate at $t_n$.
However, cavity loss without time-dependent control is also
sufficient to generate $|\Psi_{e}\rangle$ with a fidelity
approaching unity if the photon loss rates $\kappa_{1,2}$ for
modes $|\sigma_1^+\rangle$ and $|y_2\rangle$ fulfill
\begin{equation}
\kappa_1 < g/\hbar \simeq \kappa_2. \label{eq:leak-cond}
\end{equation}
In this regime, a photon in state $|y_2\rangle$ leaves the cavity
before it is scattered back into $|\sigma_1^+\rangle$, thus
terminating the time evolution in Fig.~\ref{Fig2}(a) on average
after one half-period. The condition $\kappa_1 < g/\hbar$ ensures
at least one oscillation be completed. For a quantitative
estimate, we integrate the Master equation for the density matrix
of the QD-cavity system,
\begin{equation} \dot{\hat{\rho}}(t) =
-(i/\hbar)[\hat{H},\hat{\rho}(t)]+\hat{\cal L}_{l} \hat{\rho},
\label{eq:mastereq}
\end{equation}
where cavity loss from $|\sigma_1^+\rangle$ and $|y_2\rangle$ into
free modes propagating along directions $1$ and $2$, respectively,
is described by the standard Liouville operator
\begin{equation}
\hat{\cal L}_{l} \hat{\rho} = - \sum_{i=1,2} \frac{\kappa_i}{2}
\left(\hat{a}_i^\dagger \hat{a}_i \hat{\rho} + \hat{\rho}
\hat{a}_i^\dagger \hat{a}_i - 2 \hat{a}_i^\dagger \hat{\rho}
\hat{a}_i \right).  \label{eq:mastereq2}
\end{equation}
The overall fidelity ${\cal F}$ for generation of a spin-photon
entangled state as in Eq.~(\ref{eq:ent}) is determined by the
dynamics of $\hat{\rho} (0) = | \! \uparrow; \sigma_1^+\rangle
\langle \uparrow; \sigma_1^+|$, where photon loss from mode $2$
corresponds to successful photon transfer from $1$ to $2$. The
probability for photon loss along $1$ and $2$ as a function of
time can be obtained from numerical integration of
Eq.~(\ref{eq:mastereq}) [shown in Fig.~\ref{Fig2}(b) for
$\kappa_1=0.2 g/\hbar$ and $\kappa_2 = g/\hbar$]. For
$t\rightarrow \infty$, the probability $p_2$ for photon loss into
a mode propagating along $2$ is calculated from the
Fourier-Laplace transform of Eq.~(\ref{eq:mastereq}),
\begin{eqnarray}
p_2 &= &\kappa_2 \int_0^\infty dt \, \langle \uparrow; y_2|
\rho(t) | \! \uparrow; y_2 \rangle \nonumber \\ &=& \frac{4
\kappa_2
\left(g/\hbar\right)^2}{\left(\kappa_1+\kappa_2\right)\left[4(g/\hbar)^2
+ \kappa_1 \kappa_2 \right]}.
 \label{eq:loss-prob}
\end{eqnarray}
For the parameters in Fig.~\ref{Fig2}(b), $p_2 = 79$\%. If the
photons propagate freely outside the cavity, the entanglement of
the QD spin and the photon propagation direction is preserved even
after photons are ejected from the cavity. In the regime of
Eq.~(\ref{eq:leak-cond}), the fidelity ${\cal F}=p_2$ for
generating spin-photon entanglement approaches unity.

In the ideal case $p_2 \simeq 100$\%, a (maximally entangled) Bell
state is obtained for an electron spin prepared in $(|\!
\uparrow\rangle + |\!\downarrow\rangle)/\sqrt{2}$, which evolves
according to $(|\! \uparrow;\sigma_1^+\rangle +
|\!\downarrow;\sigma_1^+\rangle)/\sqrt{2} \rightarrow (-|\!
\uparrow;y_2 \rangle + |\!\downarrow;\sigma_1^+\rangle)/\sqrt{2}
$.~\cite{rem7} We next quantify the entanglement of the final
state for a lossy cavity. As long as the photons are not detected
outside the cavity and loss in the propagating modes is
negligible, the initial state evolves into a pure
state,~\cite{enk:97,lange:00,sun:04,rem8} for which the
entanglement $E$ is given by the von Neumann entropy of the
reduced density matrix.~\cite{bennett:96,wootters:98} The
entanglement can be expressed in terms of $p_2$ in
Eq.~(\ref{eq:loss-prob}). Defining $\lambda_\pm = (1\pm
\sqrt{1-p_2})/2$, $E(p_2) = - \sum_{\sigma = \pm}\lambda_\sigma
\log_2 \lambda_\sigma$. Of particular interest are the limiting
cases of large and small $p_2$, where $\lim_{p_2 \rightarrow 1^-}
E(p_2) = 1 - (1-p_2)/4 \ln 2 + O((1-p_2)^2)$ and $\lim_{p_2
\rightarrow 0^+} E(p_2) = (1+ \ln 4 - \ln p_2)p_2/4 \ln 2 +
O(p_2^2)$, respectively. We illustrate the qualitative dependence
of $E$ on $\kappa_2$ for fixed $\kappa_1 \ll g/\hbar$. For
$g/\hbar< \kappa_2 \lesssim 4(g/\hbar)^2/\kappa_1$, loss along
direction $2$ is dominant for the spin state $|\!
\uparrow\rangle$, such that $p_2 \simeq 1$ [Fig.~\ref{Fig2}(b) and
Eq.~(\ref{eq:loss-prob})] and $E$ is of order unity. By contrast,
for large cavity loss $\kappa_2 \gtrsim 4(g/\hbar)^2/\kappa_1$,
$p_2\simeq 4(g/\hbar)^2/\kappa_1 \kappa_2$ approaches zero because
the large linewidth of $|y_2\rangle$ renders photon transfer
between the cavity modes inefficient. The entanglement $E$
decreases to zero because the photon leaves the cavity along
direction $1$ irrespective of the spin state on the QD.

Generation of spin-photon entanglement requires fine tuning of the
cavity design to ensure $g_1=g_2$ (Ref.~\onlinecite{rem4}) and
alignment of the nanocrystal. We next quantify errors for $g_1
\neq g_2$, finite detuning $\delta \neq 0$, QD misalignment, and
transitions involving lh states. In the ideal case, an initial
state $|\! \uparrow;\sigma_1^+\rangle$ evolves to $|\! \uparrow;
y_2\rangle$ with $100$\% fidelity, while ${\cal F} =  \max_t
|\langle \uparrow; y_2|\exp(-i\hat{H}t/\hbar)|\!\uparrow
;\sigma_1^+ \rangle|^2$ quantifies the fidelity for non-ideal
situations. For $g_1 \neq g_2$, ${\cal
F}=1-[(g_1^2-g_2^2)/(g_1^2+g_2^2)]^2$, which remains close to
unity for $|g_1-g_2|/|g_1+g_2| \lesssim 1/2$ [Fig.~\ref{Fig2}(c)].
A finite detuning $\delta$ of the cavity modes relative to the
hh-trion transition leads to ${\cal F} = 1 - {\cal O}(\delta/g)^2$
for $\delta \lesssim g$, which demonstrates the pivotal importance
of resonant modes. Misalignment of the QD relative to the photon
propagation directions modifies the optical selection rules. For
definiteness, consider a nanocrystal with an anisotropy axis
rotated by $\theta$ in the plane of the cavity. For $\theta \neq
0$, the coupling energy of $|\sigma_1^+ \rangle$ and transitions
from the $j_z=\pm 3/2$ hh states is $g(1\pm \cos \theta)/2$. The
dynamics of the system remain periodic for $\theta \neq 0$ and
${\cal F}=[2(1+\cos \theta)/(3+\cos^2 \theta)]^2  \simeq 1 -
\theta^4/8$ for $\theta \rightarrow 0$, i.e., the fidelity
decreases slowly for $\theta \lesssim 0.5$ [Fig.~\ref{Fig2}(d)].
Transitions involving lh states are suppressed relative to hh
processes by the small factor $g/\Delta$.

\section{Spin-photon \textsc{swap}}
\label{sec:swap}

\begin{figure}[t!]
\centerline{\mbox{\includegraphics[width=8.3cm]{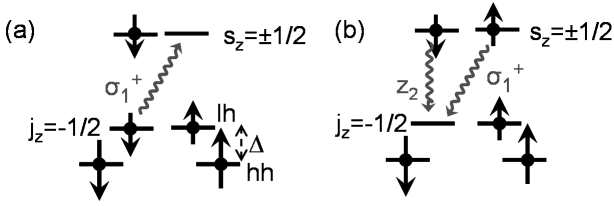}}}
\caption{(a) Absorption process involving lh valence band states.
(b) The lh-trion state $|X^-_l\rangle$ can decay by emission of a
photon into mode $|\sigma_1^+\rangle$ {\it or} $|z_2\rangle$. The
requirement that $|\sigma_1^+\rangle$ and $|z_2\rangle$ are the
only resonant modes ensures that photon emission into
$|z_2\rangle$ is accompanied by a spin flip. }\label{Fig3}
\end{figure}

We show next that, for a QD with a lh valence band maximum, the
interaction with two cavity modes allows one to implement a
\textsc{swap} gate of spin and photon states.~\cite{rem6} We
consider a cavity with the geometry shown in Fig.~\ref{Fig1}(b),
for which the circularly polarized mode $|\sigma_1^+\rangle$ and
the linearly polarized mode $|z_2\rangle$ are in resonance with
the lh-trion transition while $|y_2\rangle$ is off-resonant. While
$|\! \uparrow ;\sigma_1^+ \rangle$ is an energy eigenstate because
of Pauli blocking, the state $|\! \downarrow ;\sigma_1^+ \rangle$
exhibits dynamics similar to Eq.~(\ref{eq:ham}). Photon absorption
induces transitions to the lh-trion state $|X^-_l\rangle =
\hat{c}^\dagger_+ \hat{c}^\dagger_- \widehat{l}_-|G\rangle$, where
$\widehat{l}_\pm$ annihilates an electron in the lh state with
$j_z = \pm 1/2$ [Fig.~\ref{Fig3}(a)]. Because both
$|\sigma_1^+\rangle$ and $|z_2\rangle$ are resonant with the trion
transition, $|X^-_l\rangle$ has two different decay channels
[Fig.~\ref{Fig3}(b)]. Optical selection rules imply that, by
emission of a photon in state $|\sigma_1^+\rangle$, the QD returns
to its original spin state $|\! \downarrow \rangle$ while emission
into mode $|z_2\rangle$ leaves the QD in $|\! \uparrow\rangle$.
Hence, transfer of a photon from $|\sigma_1^+\rangle$ to
$|z_2\rangle$ is accompanied by a spin flip on the QD, which is
described by the Hamiltonian~\cite{rem1}
\begin{equation}
\hat{H} =  g_1 | X_l^-;0\rangle  \langle \downarrow;\sigma^+_1| -
g_2 | X_l^-;0\rangle  \langle \uparrow;z_2 |
 + h.c. \label{eq:ham2}
\end{equation}
with coupling constants $g_{1,2}$. The dynamics of an initial
state $|\Psi\rangle = \alpha |\uparrow;\sigma_1^+\rangle + \beta
|\downarrow; \sigma_1^+\rangle$ are readily evaluated. In
particular, for $g_1 = g_2 = g$,~\cite{rem5} we find that at time
$t_n = h (2n+1)/\sqrt{8}g$,
\begin{equation}
\alpha |\! \uparrow;\sigma_1^+\rangle + \beta
|\!\downarrow;\sigma_1^+\rangle \rightarrow  |\Psi(t_n)\rangle =
\alpha |\uparrow; \sigma_1^+ \rangle + \beta |\uparrow; z_2
\rangle, \label{eq:swap}
\end{equation}
i.e., the QD spin state is swapped onto the photon state encoded
in the amplitudes of modes $1$ and $2$, respectively. This
\textsc{swap} gate is based on optical selection rules which
enforce that photon transfer between the modes is accompanied by a
spin flip on the QD. In contrast to schemes such as in
Ref.~\onlinecite{leuenberger:04}, no additional spin measurements
are required. The reverse process of Eq.~(\ref{eq:swap}), in which
the photon state  $\alpha |\sigma_1^+\rangle + \beta |z_2\rangle$
is transferred onto a QD prepared in an initial state $|\!
\uparrow\rangle$ can also be realized by time evolution under
Eq.~(\ref{eq:ham2}). Then, $\alpha |\! \uparrow;\sigma_1^+\rangle
+ \beta |\!\uparrow;z_2\rangle \rightarrow  \alpha
|\uparrow;\sigma_1^+\rangle + \beta |\downarrow;
\sigma_1^+\rangle$. Photon states of the form $\alpha
|\sigma_1^+\rangle + \beta |z_2\rangle$, in which one photon
propagates in spatially separated modes, serve as logical basis
for linear optics quantum computing.~\cite{knill:01} The
\textsc{swap} operation in Eq.~(\ref{eq:swap}) provides a natural
interface between such coherent photon states and spins. A photon
ejected from the cavity can be converted into the standard logical
basis $\alpha |z_1\rangle + \beta |z_2\rangle$ by linear optical
elements.

Implementation of the spin-photon \textsc{swap} gate with unity
fidelity requires the interaction between photons and QD be
terminated at $t_n$. For cavity loss rates which fulfill
Eq.~(\ref{eq:leak-cond}), no time-dependent control of the cavity
parameters is required because cavity loss from mode $2$ is
sufficient to terminate the dynamics. The fidelity ${\cal F}=p_2$
derived in Eq.~(\ref{eq:loss-prob}) approaches unity. For
$\kappa_1=0.2g/\hbar$ and $\kappa_2=g/\hbar$, ${\cal F}= 79$\%.

\section{Generation of entangled photon pairs}
\label{sec:photons}

\begin{figure}
\centerline{\mbox{\includegraphics[width=8.3cm]{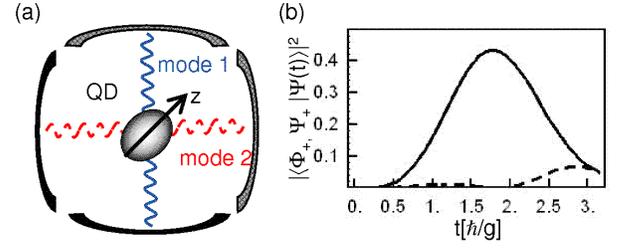}}}
\caption{(a) Setup for the generation of photon-entanglement by
strong-coupling dynamics. (b) Projection of $|\Psi(t)\rangle =
e^{-i\hat{H}t/\hbar}|\sigma_1^+\rangle | \sigma_1^-\rangle \otimes
|G\rangle$ onto the Bell states $|\Psi_+\rangle$ (solid line) and
$|\Phi_+\rangle$ (dashed line).}\label{Fig4}
\end{figure}

Similarly to an atom coupled to two cavity
modes,~\cite{wildfeuer:03} an {\it undoped} QD with symmetry axis
$z$ at $45^\circ$ relative to the propagation directions of modes
$1$ and $2$ [Fig.~\ref{Fig4}(a)] acts as entangler for photon
pairs. We consider a QD with the level scheme in
Fig.~\ref{Fig1}(a) and assume that the four photon states
$|\sigma_{1,2}^\pm\rangle$ are resonant with the lowest (hh)
exciton transition. Absorption of photons from mode $1$ and
re-emission into $2$ generates polarization entangled pairs from
an initial product state $|\sigma_1^+\rangle| \sigma_1^-\rangle$.
The interaction strength of $|\sigma_{1,2}^+\rangle$ and the
exciton states $|X_\pm\rangle = \hat{c}^\dagger_\mp \hat{h}_\mp
|G\rangle$ is parameterized by the coupling constants $g(1\pm
1/\sqrt{2})/2$. Because transitions from
$|\sigma_{1,2}^\pm\rangle$ to $|X_{\pm}\rangle$ are dominant, for
short times $t < h/g$ an initial state prepared by injecting a
photon pair $|\sigma_1^+\rangle| \sigma_1^-\rangle$ into the
cavity evolves predominantly according to the sequence
$|\sigma_1^+\rangle | \sigma_1^-\rangle \otimes |G\rangle
\rightarrow \left(|\sigma_1^+\rangle \otimes |X_-\rangle +
|\sigma_1^-\rangle \otimes |X_+\rangle \right)/\sqrt{2}
\rightarrow \left(|\sigma_1^+\rangle | \sigma_2^-\rangle +
|\sigma_1^- \rangle |\sigma_2^+\rangle \right)\otimes
|G\rangle/\sqrt{2}$.

Rigorously, all allowed transitions must be taken into account, $
\hat{H} =  g \sum_{\alpha,\beta=\pm; i=1,2} u_{\alpha \cdot \beta}
\hat{a}_{i,\alpha} |X_{\beta} \rangle \langle G| + h.c.$, where
$u_\pm = (1 \pm 1/\sqrt{2})/2$ and $\hat{a}_{i,\pm}$ annihilates a
photon in state $|\sigma_i^\pm \rangle$. Because $g$ is small
compared to the biexciton shift, biexciton states can be
neglected. Integrating the Schr{\"o}dinger equation, we obtain
\begin{eqnarray} && |\Psi(t)
\rangle = - \frac{1- \cos (2 u_+ g t/\hbar) \cos
(2 u_- g t/\hbar)}{2 \sqrt{2}}|\Psi_{+}\rangle \nonumber \\
 && \hspace*{0.5cm}- \frac{\sin (2 u_+ g t/\hbar)\sin
(2 u_- g t/\hbar) }{2\sqrt{2}} |\Phi_{+}\rangle +
|\tilde{\Psi}\rangle \label{eq:entangler}
\end{eqnarray}
with the Bell states $|\Psi_{+}\rangle = \left(|\sigma_1^+\rangle
|\sigma_2^-\rangle + |\sigma_1^- \rangle |\sigma_2^+\rangle
\right)\otimes |G\rangle/\sqrt{2}$ and $|\Phi_{+}\rangle =
\left(|\sigma_1^+\rangle |\sigma_2^+\rangle + |\sigma_1^- \rangle
|\sigma_2^-\rangle \right)\otimes |G\rangle/\sqrt{2}$.
$|\tilde{\Psi}\rangle$ represents components with zero photons in
one of the modes. Figure~\ref{Fig4} shows the projection onto
$|\Psi_{+}\rangle$ (solid line) and $|\Phi_{+}\rangle$ (dashed
line) as a function of time. As expected, for $t \lesssim h/g$ the
transition to the polarization entangled state $|\Psi_+\rangle$ is
dominant. At $t_n = h n/4 u_{\pm}g$, the $|\Phi_+\rangle$
component vanishes. While instantaneous reduction of the cavity
$Q$-factor at $t_n$ would allow one to extract $|\Psi_+\rangle$
from the cavity with a fidelity limited only by off-resonant
transitions, cavity loss rates $\kappa_{1,2} \simeq g/\hbar$ are
also sufficient to terminate the coherent dynamics in
Fig.~\ref{Fig4}(b). Hence, an undoped QD strongly coupled to
several modes of a lossy cavity acts as efficient entangler of
photon pairs.

\section{Discussion of experimental parameters}
\label{sec:exp}

While our calculations in Secs.~\ref{sec:entanglement},
\ref{sec:swap}, and \ref{sec:photons} show that a QD interacting
with two cavity modes has interesting applications as interface
between spin and photon quantum states, the system is difficult to
implement experimentally. Cavities based on Bragg reflectors can
sustain degenerate circularly and linearly polarized modes, but
mode volumes of order $\lambda^3$ are impossible to reach because
of diffraction. We show next how the two-mode Jaynes-Cummings
Hamiltonian in Eq.~(\ref{eq:ham}) [Fig.~\ref{Fig1}(b)] can in
principle be implemented with optical microcavities, where small
mode volumes  can be achieved. Because the coupling constants in
Eq.~(\ref{eq:ham}) are determined by the electric fields at the
site of the QD only, it is sufficient that the mode
$|\sigma_1^+\rangle$ is circularly polarized {\it locally}, at the
site of the QD. For definiteness, we focus on the defect modes in
a triangular photonic crystal, with a central hole (the defect)
with radius $r_{d}$ and dielectric constant $\epsilon_d$ which is
different from that of all other holes in the triangular lattice.
The defect modes with electric field in the cavity plane (TM) and
perpendicular to the cavity plane (TE) have been analyzed in
detail for some specific realizations of the background and hole
medium.~\cite{kuzmiak:98,kuzmiak:00,stojic:01} The defect mode
energies are proportional to $r_d/\sqrt{\epsilon_d}$ and can be
tuned across the optical bandgap by varying $r_d$ and
$\epsilon_d$.~\cite{kuzmiak:98,kuzmiak:00,stojic:01,rem9}

The following steps allow one to experimentally implement the
two-mode Jaynes-Cummings model in Eq.~(\ref{eq:ham}): (i) Choose
$\epsilon_d$ and $r_d$ such that a doubly degenerate TE mode
(e.g., the $E_1$ or $E_2$ mode~\cite{kuzmiak:00}) is degenerate
with one TM mode. For a triangular photonic crystal with hexagonal
holes, the coexistence of degenerate TE and TM defect modes has
recently been demonstrated.~\cite{matthews:03} We refer to the
modes of the TE-doublet as $|{\rm TE}_{1/2}\rangle$ and to the TM
mode as $|{\rm TM}\rangle$. $|{\rm TE}_1\rangle$ and $|{\rm
TE}_2\rangle$ are related by a $\pi/2$-rotation.~\cite{kuzmiak:00}
(ii) Identify the set of points $\{P\}=\{(x,y)|E_{|{\rm
TE}_{1}\rangle}=E_{|{\rm TM}\rangle}\}$ in the cavity plane where
the electric field amplitudes $E_{|{\rm TE}_{1}\rangle}$ and
$E_{|{\rm TM}\rangle}$ of $|{\rm TE}_1\rangle$ and $|{\rm
TM}\rangle$ are equal. The points $\{P\}$ typically form a set of
several lines. For every point in $\{P\}$, $|\sigma_1^+\rangle =
(|{\rm TE}_1\rangle + i |{\rm TM}\rangle)/\sqrt{2}$ locally
generates an electric field with circular polarization. (iii) In
$\{P\}$, identify a point $(x_{\rm QD},y_{\rm QD})$ where the
electric field amplitude $E_{|{\rm TE}_{2}\rangle}$ of $|{\rm
TE}_2\rangle$ lies within $30$\% of $E_{|{\rm
TE}_{1}\rangle}/\sqrt{2}$.~\cite{rem10} For a QD at $(x_{\rm
QD},y_{\rm QD})$ with anisotropy axis oriented perpendicular to
the electric field of $|{\rm TM}\rangle$ in the cavity plane, the
QD-cavity interaction is described by the Hamiltonian
Eq.~(\ref{eq:ham}) with $|g_2/g_1-1| \leq 0.3$, which guarantees a
theoretical fidelity of at least $90$\% [Fig.\ref{Fig2}(c)]. Note
that high cavity Q-factors are maintained for a wide range of
$\epsilon_d$ and $r_d$.~\cite{kuzmiak:98,kuzmiak:00,stojic:01}

Additional requirements for the dynamics discussed in
Secs.~\ref{sec:entanglement} and \ref{sec:swap} include a cavity
loss rate $\kappa_2$ large compared to $\kappa_1$ and the
injection of a single photon into $|\sigma_1^+\rangle$. Because
$|{\rm TE}_2\rangle$ is predominantly localized along one
direction of the photonic crystal,~\cite{kuzmiak:00} the
corresponding cavity loss rate $\kappa_2$ can be increased by
reducing the size of the photonic crystal in this direction, i.e.,
by removing holes at the outside. While this changes the energies
of all three modes, $|{\rm TE}_{1,2}\rangle$ and $|{\rm
TM}\rangle$, the energy shifts are negligible for cavities with
large $Q$-factors. Injection of a single photon into mode
$|\sigma_1^+\rangle$ can be achieved by irradiation of the
microcavity with a single-photon source. For TE defect modes in
small cubic photonic crystals, the injection efficiency was
calculated to be of order $50$\%.~\cite{villeneuve:96} Photon
injection into $|\sigma_1^+\rangle = (|{\rm TE}_1\rangle + i |{\rm
TM}\rangle)/\sqrt{2}$ is more complicated because coupling
efficiencies can be different for TE and TM modes and depend on
the direction of incidence relative to the photonic crystal. One
can overcome this problem by determining the directions for which
the coupling efficiencies for $|{\rm TE}_1\rangle$ and $|{\rm
TM}\rangle$ are comparable, using numerical techniques similar to
those in Ref.~\onlinecite{villeneuve:96}. Alternatively, a source
of elliptically polarized photons can be used, where the TE and TM
field amplitudes compensate the difference in coupling
efficiencies. We also note that a  high photon injection
efficiency is not required as long as unsuccessful injection
attempts can be excluded by post-selection.

For a quantitative estimate, we consider spherical CdSe
nanocrystals with a mean radius $a=5$~nm. The energy of the lowest
exciton state $1S_{3/2}$-$1S_e$ in an undoped QD is $E_{X}=1.93 \,
{\rm eV}$ (Refs.~\onlinecite{efros:92,ekimov:93}) while the trion
transition is redshifted by $0.5$~meV. The hh-lh splitting of a
spherical QD, $\Delta \simeq 20~{\rm meV}$, is large compared to
the coupling constant $g$. For a mode volume $(\lambda/n)^3$, with
$n$ the refractive index of the cavity, the electric field
amplitude of the cavity modes is of order
$E=\sqrt{2E_{X}n/\epsilon_0 \lambda^3}=5\times 10^5 \sqrt{n} \,
\,{\rm V}/{\rm m}$. With the Kane interband matrix element
$\langle S|\hat{p}_y|Y \rangle$,~\cite{ekimov:93} $g=(eE/m
\omega)|\langle S|\hat{p}_y|Y \rangle |\simeq 0.2~{\rm meV}$.
Strong-coupling phenomena require $g$ to be large compared to both
the spontaneous QD emission rate and the cavity loss rates
$\kappa_{1,2}$. PL linewidths of $0.12 \,{\rm meV} < g$ have been
observed for individual CdSe nanocrystals.~\cite{empedocles:96}
For cavity $Q$-factors of order $10^4$, $\kappa = \omega/Q
\lesssim g/h$. In addition, the phenomena discussed here require a
hole spin relaxation time long compared to $h/g \simeq 20~{\rm
ps}$. Recent PL studies of CdSe QD's suggest that hole spin
relaxation times are of order $10\, {\rm
ns}$.~\cite{flissikowski:03} These values show that the
strong-coupling dynamics discussed above is within experimental
reach for CdSe nanocrystals in a microcavity. The main challenge
is to design microcavities with two modes with different
polarization, spatial distribution, and loss rates which are
strongly coupled to a QD. As shown here, this system would allow
one to generate spin-photon entanglement, implement a spin-photon
\textsc{swap} gate, and create polarization entangled photon
states.

\acknowledgements We acknowledge helpful discussions with V.
Cerletti, R.J. Epstein, S. Ghosh, O. Gywat, Y. Li, and F. Mendoza.
This work was supported by ONR and DARPA.


\begin{thebibliography}{38}
\bibitem[{rem({\natexlab{a}})}]{rem1}
\bibinfo{note}{Terms which are irrelevant for an initial photon state
  $|\sigma_1^+\rangle$ are omitted.}
\bibitem[{rem({\natexlab{b}})}]{rem7}
\bibinfo{note}{For simplicity, we denote photons propagating along directions
  $1$ and $2$ outside the cavity with the same symbols as the corresponding
  cavity modes. It is understood that, for $t\rightarrow \infty$, photons
  always occupy propagating modes outside the cavity.}
\bibitem[{rem({\natexlab{c}})}]{rem8}
\bibinfo{note}{Up to two phase factors ($\phi_1$ and $\phi_2$) which are
  irrelevant for the following quantification of entanglement, the state can be
  expressed in terms of the loss probability $p_2$ derived in
  Eq.~(\ref{eq:loss-prob}): $(|\! \uparrow;\sigma_1^+\rangle +
  |\!\downarrow;\sigma_1^+\rangle)/\sqrt{2} \rightarrow (\sqrt{1-p_2} \, e^{i
  \phi_1}|\! \uparrow;\sigma_1^+ \rangle + \sqrt{p_2} \, e^{i \phi_2} |\!
  \uparrow;y_2 \rangle + |\!\downarrow;\sigma_1^+\rangle)/\sqrt{2}$.}
\bibitem[{rem({\natexlab{d}})}]{rem4}
\bibinfo{note}{$g_1=g_2$ if $E_1/E_2=1/\sqrt{2}$, where $E_{1,2}$ is the
  electric field amplitude of one photon in $|\sigma_1^+\rangle$ or
  $|y_2\rangle$, respectively.}
\bibitem[{rem({\natexlab{f}})}]{rem5}
\bibinfo{note}{$g_1=g_2$ if $E_1/E_2=\sqrt{2}$, where $E_{1,2}$ is the electric
  field amplitude of one photon in $|\sigma_1^+\rangle$ or $|z_2\rangle$,
  respectively.}
\bibitem[{rem({\natexlab{g}})}]{rem9}
\bibinfo{note}{The detailed dependence of the mode energies on $r_d$ and
  $\epsilon_d$ is typically different for TE and TM modes.}
\bibitem[{rem({\natexlab{h}})}]{rem10}
\bibinfo{note}{While no such point is guaranteed to exist, the requirement on
  the electric field strength of $E_{|{\rm TE}_2\rangle}$ is not very
  stringent. Note also that, in addition to $r_d$ and $\epsilon_d$, at least
  two additional parameters can be varied to find such a point, namely the
  radius of the holes in the triangular photonic crystal and the dielectric
  constants of the background material.}
\end{thebibliography}

\end{document}